\documentclass[preprint,12pt]{elsarticle}

\usepackage{amssymb}
\usepackage{amsmath}

\usepackage{url}

\begin{document}

\begin{frontmatter}



\title{Flux-tube structure in finite temperature QCD}


\author[phys-uw]{M.~Baker} 
\ead{mbaker4@uw.edu}

\author[infn-bari]{P.~Cea}
\ead{paolo.cea@ba.infn.it}

\author[itp-gu]{V.~Chelnokov}
\ead{chelnokov@itp.uni-frankfurt.de}

\author[infn-bari]{L.~Cosmai}
\ead{leonardo.cosmai@ba.infn.it}

\author[infn-cosenza,phys-unical]{A.~Papa}
\ead{alessandro.papa@fis.unical.it}

\affiliation[phys-uw]{
    organization={Department of Physics, University of Washington},
    city={Seattle},
    postcode={WA 98105},
    country={USA}}

\affiliation[infn-bari]{
    organization={INFN - Sezione di Bari},
    city={Bari},
    postcode={I-70126},
    country={Italy}}

\affiliation[itp-gu]{
    organization={Institut f\"ur Theoretische Physik, Goethe Universit\"at}, 
    city={\allowbreak Frankfurt~am~Main}, 
    postcode={60438}, 
    country={Germany}}

\affiliation[infn-cosenza]{
    organization={INFN - Gruppo collegato di Cosenza},
    city={Arcavacata di Rende, Cosenza},
    postcode={I-87036},
    country={Italy}}

\affiliation[phys-unical]{
    organization={Dipartimento di Fisica dell’Università della Calabria},
    city={\allowbreak Arcavacata~di~Rende~(CS)},
    postcode={87036},
    country={Italy}}

\begin{abstract}
We present a study of the structure of the chromoelectrical field created by a static quark-antiquark pair in lattice QCD with 2+1 flavours of dynamical quarks, where the quark masses are set to their physical values. The analysis covers a wide range of temperatures both above and below the chiral crossover, and explores varying quark-antiquark distances, with the aim of identifying signals of deconfinement and string breaking in the field structure. To this end we apply the zero-curl perturbative field subtraction method, developed in our earlier studies of pure gauge SU(3) theory and of full QCD at zero temperature.
\end{abstract}


\end{frontmatter}



\section{Introduction}
\label{sec:intro}
Quark confinement is a fundamental feature of QCD that
still lacks a complete theoretical explanation from first principles. Phenomenologically, confinement manifests in the absense of free color charges: quarks and gluons are never observed in isolation, but only within color-neutral bound states such as mesons and baryons. 

A widely used theoretical picture of confinement 
considers infinitely heavy 
``static'' quarks in pure gauge theory, coresponding to the limit  $m_q \to \infty$. Introducing a static quark-antiquark pair at separation $d$ to the system increases its energy by an amount $V(d)$, known as the static quark potential.
At low temperatures this potential is well described by 
\begin{equation}
V(d) = -\frac{c}{d} + \sigma d \ ,
\end{equation}
with a short-distance Coulomb term and a large-distance linear term.
The energy $V(d)$ comes from the energy of chromoelectric field generated by the static quarks -- in particular, a linear term comes from a longitudinal field structure of constant cross section, called a ``flux tube'', while the Coulomb term comes from the perturbative chromoelectric field with zero curl. 

Raising the temperature in pure gauge theory leads to a first order deconfinement transition. Above this point the string tension vanishes and the Coulomb field becomes Debye-screened.

In full QCD with dynamical light quarks the picture changes --- the static quark potential does not grow indefinitely, but levels off at a finite distance $d_\mathrm{sb}$, beyond which separating the sources costs no additional energy. This ``string breaking'' reflects the screening of the 
static color charges by dynamical quark-antiquark pairs. 
Below $d_\mathrm{sb}$, however the potential still resembles that of the pure gauge theory, motivating a closer investigation of the ``flux tubes'' that generate the potential in full QCD. 

Finally, unlike the pure gauge case, full QCD features no true deconfinement transition, but a smooth crossover associated with  chiral symmetry restoration. A central goal of this work is to explore how flux tubes at intermediate separations evolve across this transition region, and whether the onset of chiral symmetry restoration coincides with the disappearance of the flux-tube-like structures and their contribution to confinement.

This work continues our investigation of chromoelectric field profiles generated by static color charges, both in pure gauge SU(3) theory \cite{su3-1,su3-2,su3-3,su3-4} and in QCD with dynamical quarks \cite{qcd-1}. We briefly summarize the key elements of our approach:
\begin{itemize}
\item We probe the fields using a connected correlator between a Wilson loop $W$ representing the static sources, and a plaquette $U_P$ as a field probe, connected by a Schwinger line $L$ (see \cite{ref-connected-correlator}):
\begin{equation}
    \rho^{\text{conn}}_{W,\mu\nu} = 
    \frac{\langle \mathrm{tr}(W L U_P L^\dagger)\rangle}{\langle \mathrm{tr}(W)\rangle}
    - \frac{1}{N}\,\frac{\langle \mathrm{tr}(U_P)\,\mathrm{tr}(W)\rangle}{\langle \mathrm{tr}(W)\rangle}\;,
    \label{connected1}
\end{equation}
In this work we use correlators with the Schwinger line attached either to the quark, or to the antiquark (see Fig.~\ref{figure:connected-operator}), which reduces the 
detrimental effect of long Schwinger lines on the field estimations near the antiquark.

\begin{figure*}[htb]
\begin{center}
\includegraphics[width=0.47\linewidth,clip]{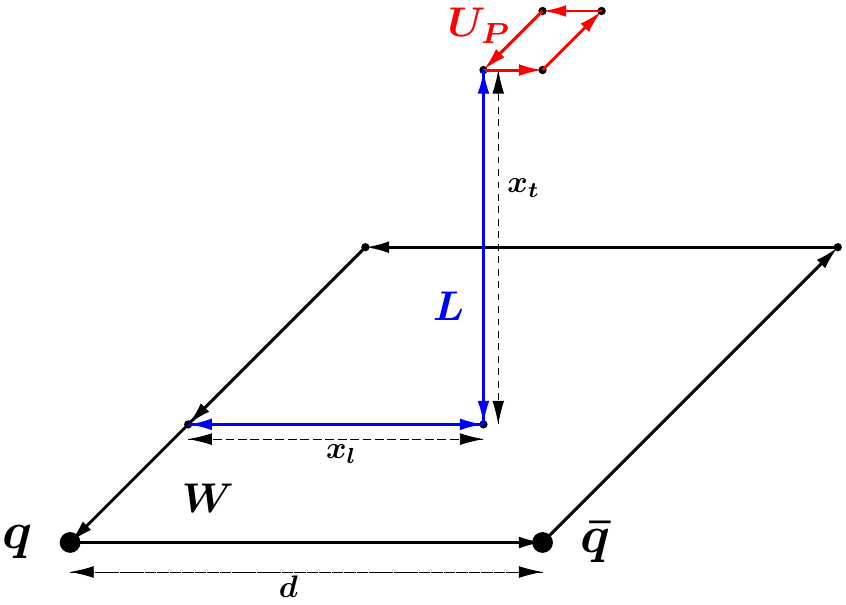}
\includegraphics[width=0.47\linewidth,clip]{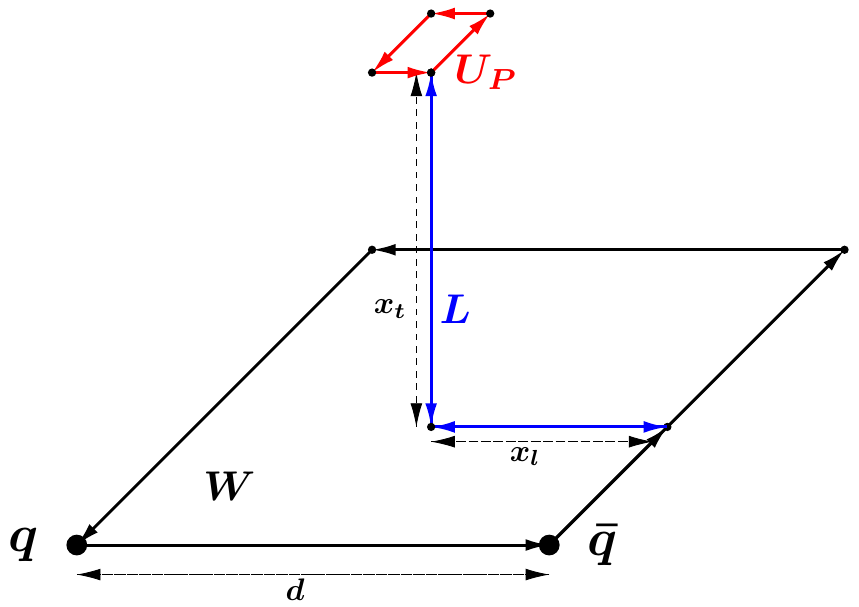}
\end{center}
\caption{Connected field operator with the Schwinger line attached to the quark (left) or antiquark (right).
}            
\label{figure:connected-operator}
\end{figure*}

The correlator $\rho^{\text{conn}}_{W,\mu\nu}$ defines a field strength tensor $F_{\mu\nu}$ via
\begin{equation}
    \rho^{\text{conn}}_{W,\mu\nu} \;\equiv\; a^2 g\,F_{\mu\nu}\;,
    \label{connected2}
\end{equation}
with different plaquette orientations used to sample different field components. 

\item Exploiting a cylindrical symmetry of the system, we probe 
the field at longitudinal distance $x_l$ along the quark-antiquark axis, and transverse distance $x_t$ perpendicular to the axis. 

\item Observables are measured on smeared gauge field configurations, which reduces stochastic noise and implements  effective multiplicative renormalization. 

\item To isolate the flux tube contribution, we apply a ``curl subtraction'' procedure. Assuming that the nonperturbative field is purely longitudinal, we equate the transverse component of the full field to that of the perturbative field, restore the longitudinal perturbative component using the zero curl condition (assuming neglible perturbative field at large transverse distances), and subtract it from the full field. This procedure removes short range effects of the Coulomb-like term and yields a reliable extraction of the flux tube profile at intermediate separations, where the perturbative field contributions can be significant compared to the nonperturbative field. 

\end{itemize}

\section{Lattice setup and numerical results}
\label{sec:lattice-setup-numerical-results}

\begin{figure*}[htb]
\begin{center}
\includegraphics[trim=-8.685 -2.111 -1.903 3.625,width=0.47\textwidth,clip]{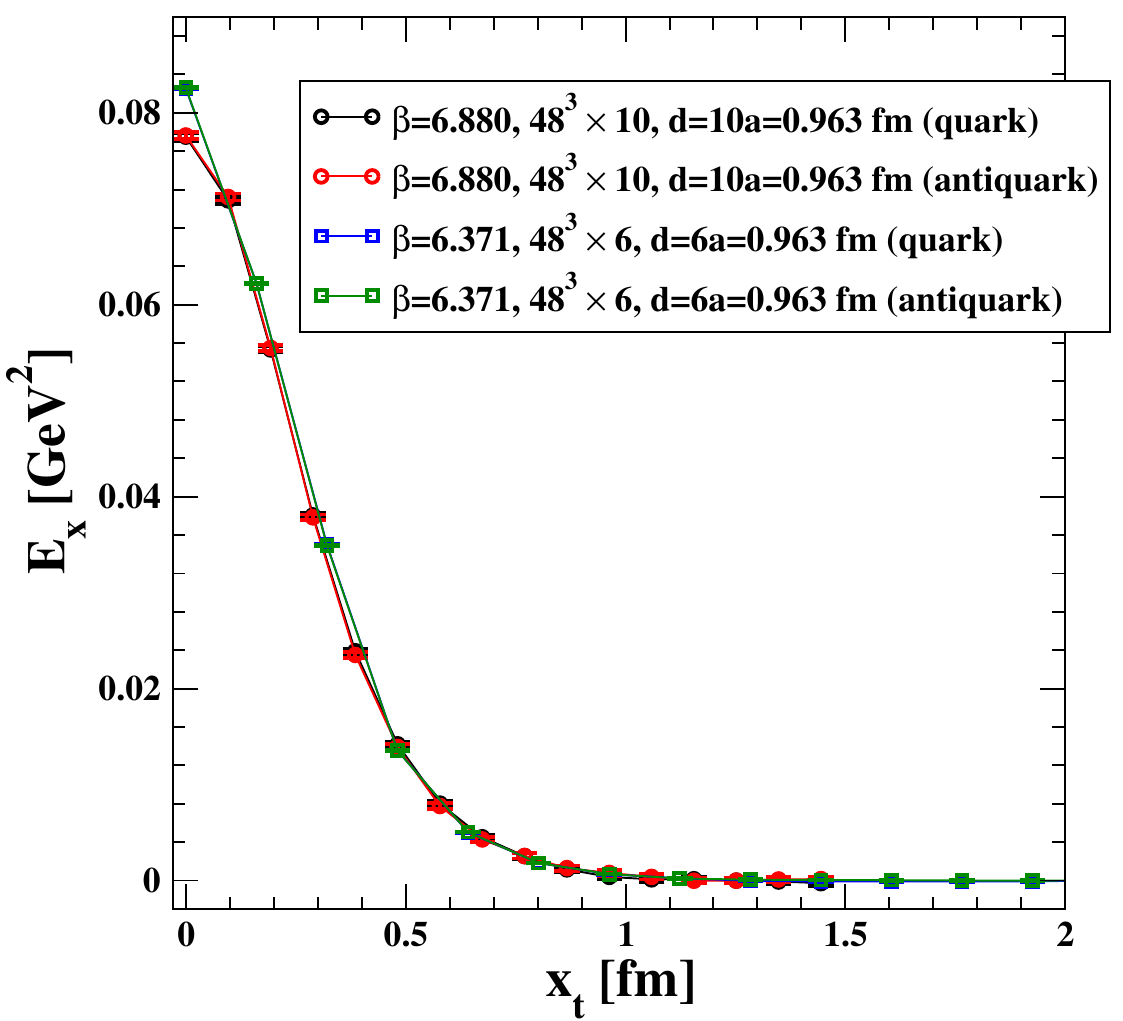}
\includegraphics[width=0.47\textwidth,clip]{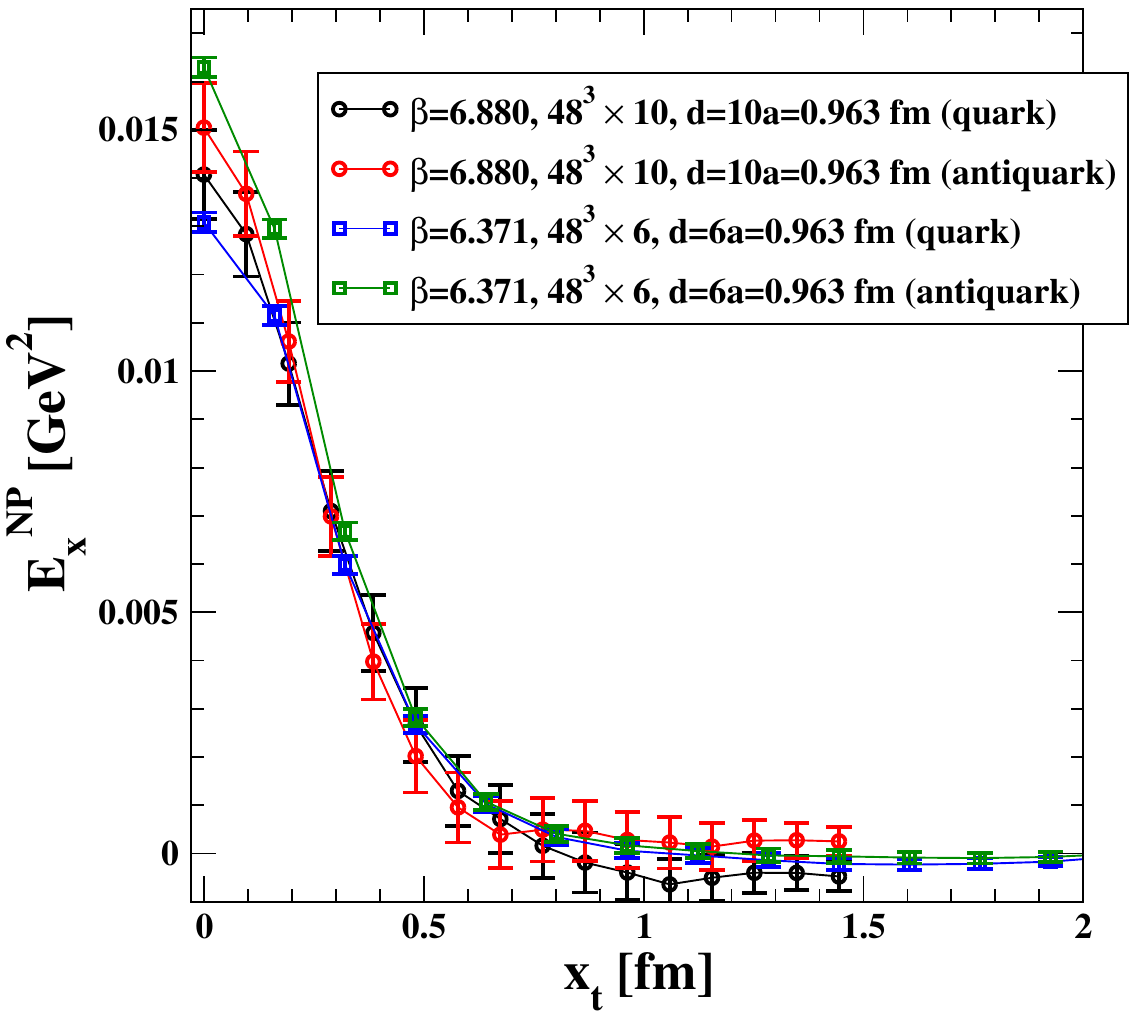}
\end{center}
\caption{Scaling check at $T \simeq 205$ MeV and $d \simeq 0.963$ fm for the full longitudinal chromoelectric field $E_x$ (left) and its nonperturbative component $E_x^{\rm NP}$ (right).
}            
\label{figure:scaling-check}
\end{figure*}

We perform 2+1 flavor lattice QCD simulations using the HISQ/tree action~\cite{Follana:2006rc,Bazavov:2009bb,Bazavov:2010ru}
along the line of constant physics, as defined in~\cite{Bazavov:2011nk}, 
with the strange quark mass fixed to its physical value and a light-to-strange mass ratio $m_l/m_s=1/27$, corresponding to a pion mass of 140 MeV.
Simulations were carried out on $48^3\times L_t$ lattices ($4 \le L_t \le 48$) with 
measurements separated by 25 trajectories of RHMC updates.

We employ one step of four-dimensional hypercubic smearing (HYPt) on the temporal links $(\alpha_1,\alpha_2,\alpha_3) = (1.0, 1.0, 0.5)$~\cite{Hasenfratz:2001hp}, followed by $N_{\rm HYP3d}$ steps of hypercubic smearing restricted to the three spatial directions (HYP3d) with parameters $(\alpha_1^{\text{HYP3d}},\alpha_3^{\text{HYP3d}}) = (0.75, 0.3)$. 
We find that $N_{\rm HYP3d} = 50$ suffice to stabilize the field components.
Figure~\ref{figure:scaling-check} shows a scaling check at $T \approx 205$ MeV and $d \approx 0.963$ fm, 
confirming that our smearing strategy is reliable and that the lattice discretization effects are under control.

\begin{figure*}[htb!]
    \begin{center}
    \includegraphics[width=0.49\textwidth,trim={1.3cm 0 1.8cm 1cm},clip]{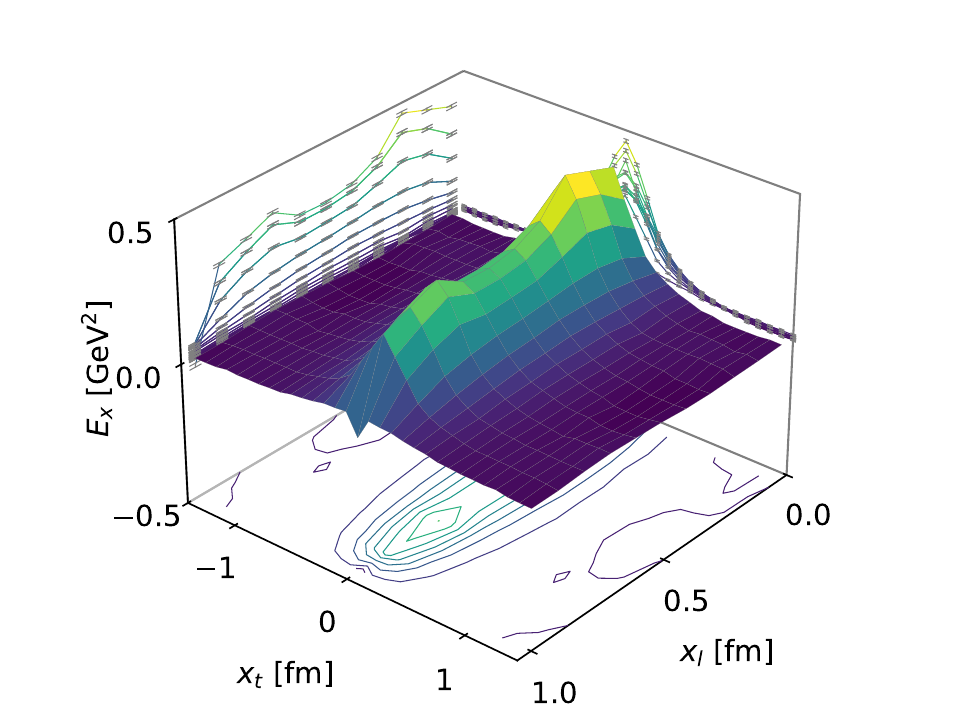}
    \includegraphics[width=0.49\textwidth,trim={1.3cm 0 1.8cm 1cm},clip]{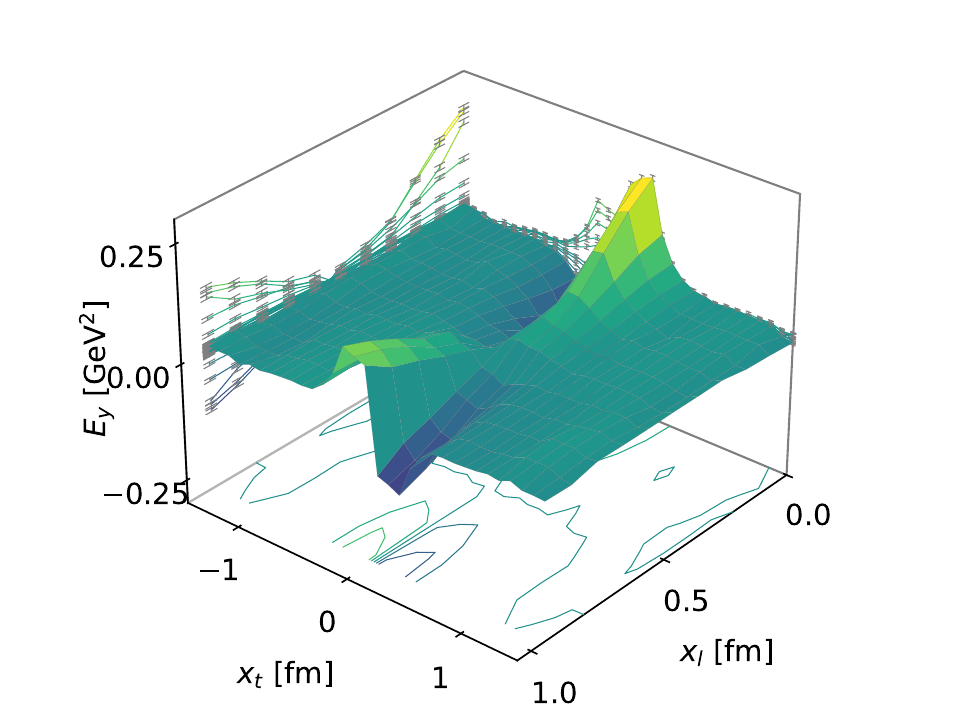} \\
    \includegraphics[width=0.49\textwidth,trim={1.3cm 0 1.8cm 1cm},clip]{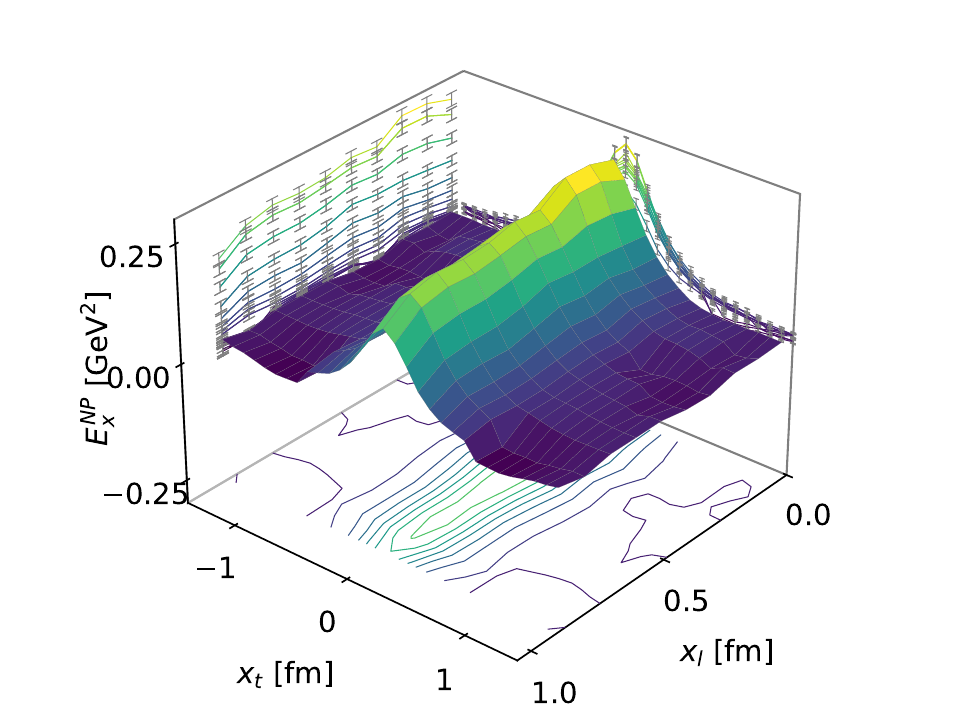}
    \includegraphics[width=0.49\textwidth]{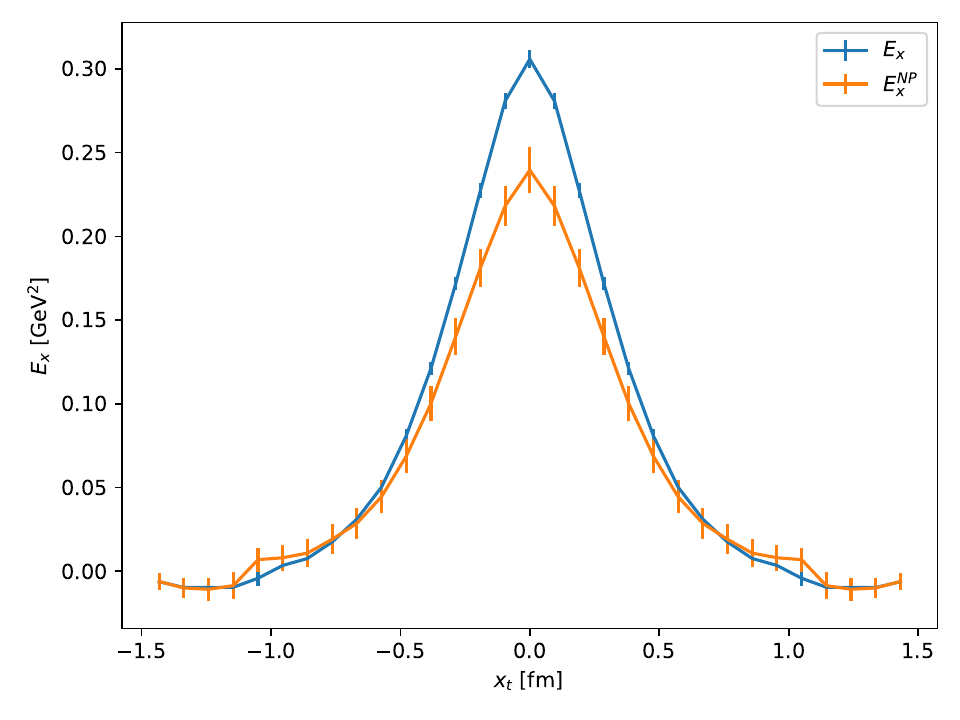}
    \end{center}
\caption{The fields at $T \approx 43$ MeV, $d \approx 0.963$ fm: 
(top left) full longitudinal field profile, 
(top right) full transverse field profile,
(bottom left) nonperturbative field profile, 
(bottom right) comparison of full and nonperturbative fields at midplane.
}            
\label{fig:plots_T43}
\end{figure*}  

\begin{figure*}[htb]
    \begin{center}
    \includegraphics[width=0.49\textwidth,trim={1.3cm 0 1.8cm 1cm},clip]{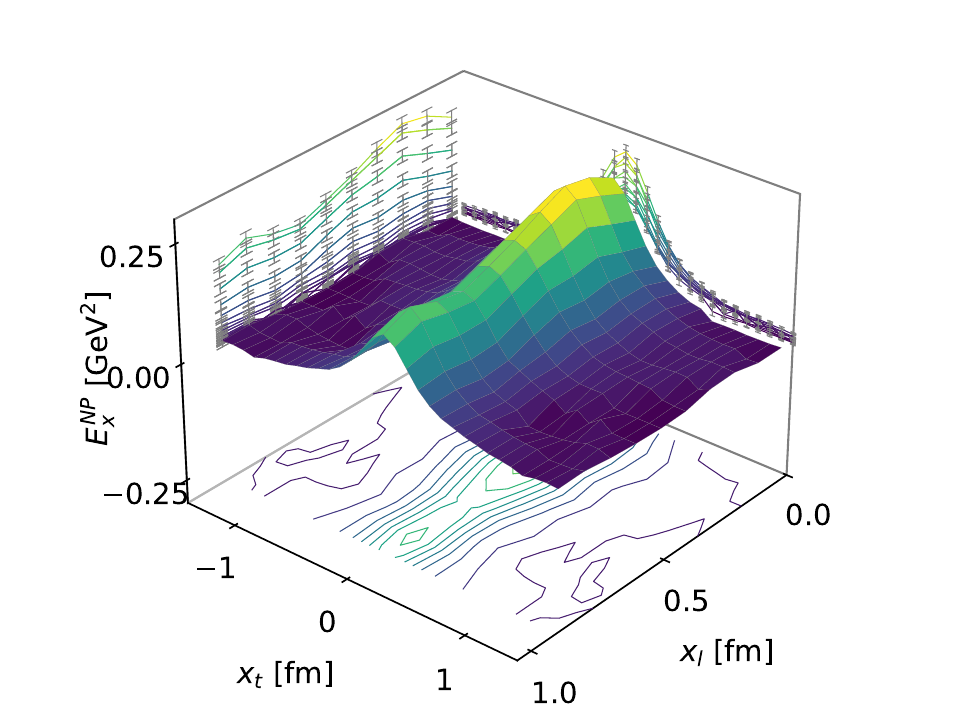}
    \includegraphics[width=0.49\textwidth]{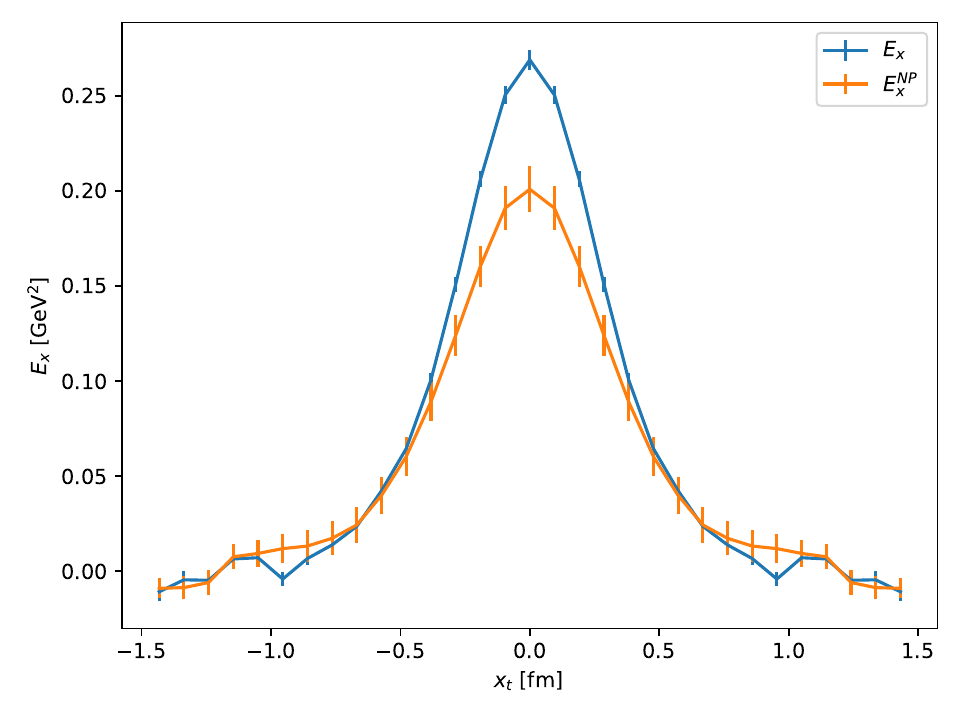}
    \end{center}
\caption{The fields at $T \approx 146$ MeV, $d \approx 0.963$ fm: 
(left) nonperturbative field profile, 
(right) comparison of full and nonperturbative fields at midplane.
}            
\label{fig:plots_T146}
\end{figure*}

\begin{figure*}[htb]
    \begin{center}
    \includegraphics[width=0.49\textwidth,trim={1.3cm 0 1.8cm 1cm},clip]{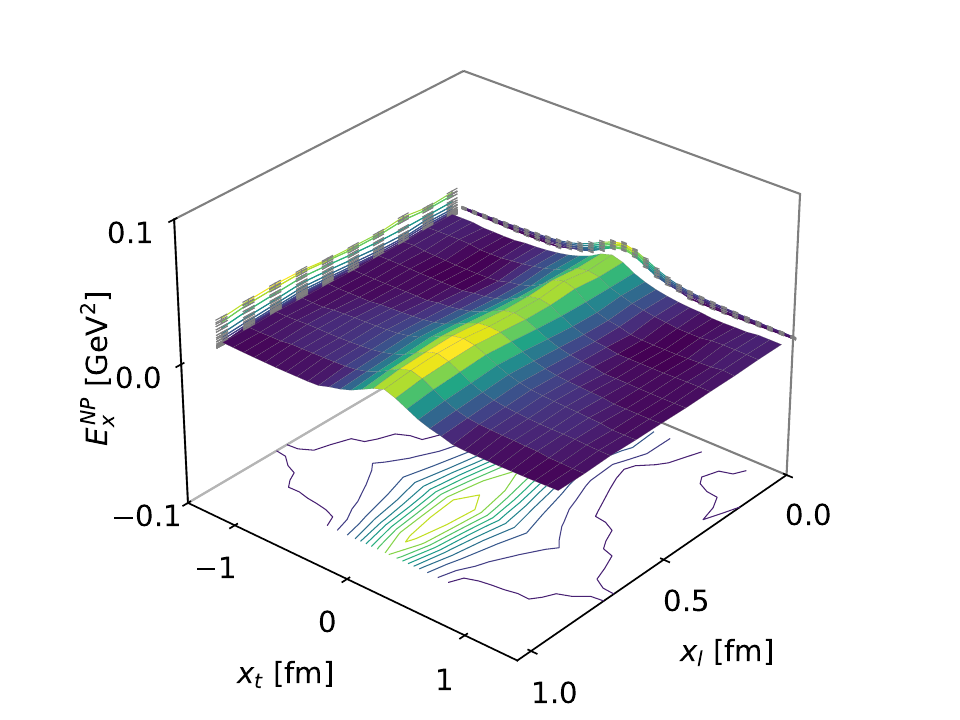}
    \includegraphics[width=0.49\textwidth]{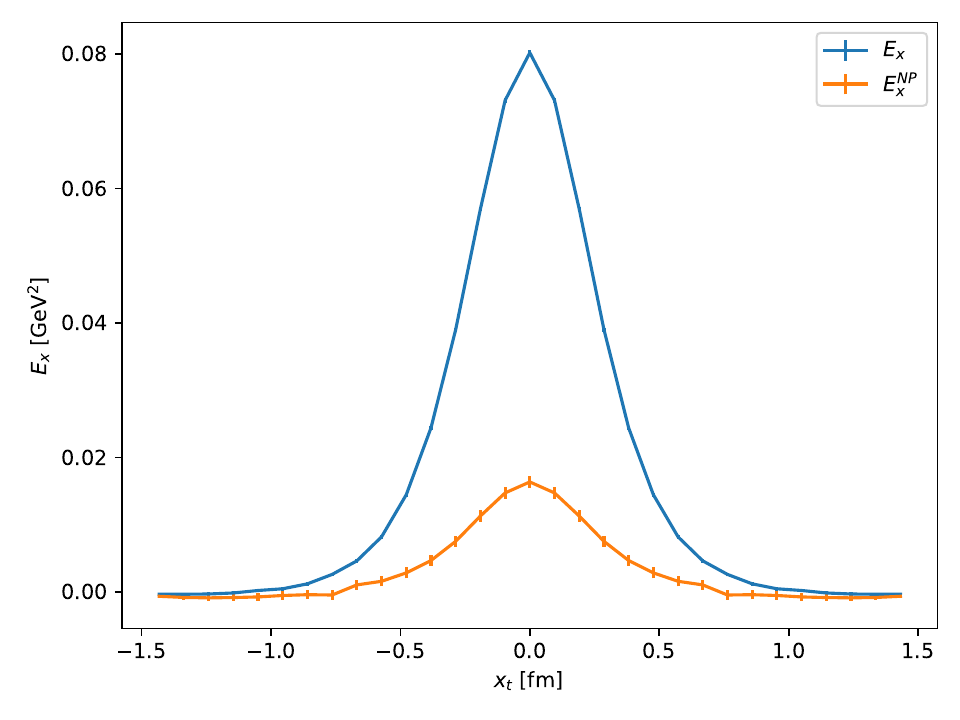}
    \end{center}
\caption{The fields at $T \approx 205$ MeV, $d \approx 0.963$ fm: 
(left) nonperturbative field profile, 
(right) comparison of full and nonperturbative fields at midplane.
}            
\label{fig:plots_T205}
\end{figure*}

Implementing the approach described in Section~\ref{sec:intro}, for our simulations and extracting the field components for different values of temperature we can see a well defined flux-tube profile both at low temperatures (Figure~\ref{fig:plots_T43}), 
temperatures around the chiral phase transition (Figure~\ref{fig:plots_T146}) and above it (Figure~\ref{fig:plots_T205}). 
The remnant flux tube structure, albeit quite weak, is still visible at $T \approx 512$ MeV (the largest temperature in our simulations), where 
$\frac{T}{T_\mathrm{pc}} \approx 3.3$. 

To quantitatively describe the flux-tube behavior across the deconfinement region we extract the effective string tension $\sigma_{\text{eff}}$ and the flux-tube width $w$
from the profiles at midpoint:
\begin{equation}
\sigma_{\text{eff}} = \int d^2 x_t \, \frac{(E_x^{\rm NP}(d/2,x_t))^2}{2} \ ,
\label{sigma_eff}
\end{equation}

\begin{equation} 
w^2 = \frac{\int d^2 x_t \, x_t^2 \, E_x^{\rm NP}(d/2,x_t)}{\int d^2 x_t \,E_x^{\rm NP}(d/2,x_t))} \ .
\label{width}
\end{equation}

We see that at small temperatures both string tension and width remain stable under changes of $T$, 
while at higher temperatures they both decrease when either $T$ increases.
To model this behavior we suggest a dependence 
\begin{equation}
\label{6.12}
\sqrt{\sigma_{\rm eff}}(d,T) \; = \;  \sqrt{\sigma_{\rm eff}}(0) \;  \exp[- \frac{1}{2} \, \mu_{\rm st}(T) \, d]   \;  \; ,
\end{equation}
where $\sqrt{\sigma_{\rm eff}}(0)$ is the zero-temperature effective string tension and
\begin{equation}
\label{6.13}
 \mu_{\rm st}(T) \; \simeq \;  
 \left\{ \begin{array}{ll}
 \; \;   0  \; \; &   \; \; T  \; \lesssim \; T_0  \; ,
  \\
 \;  \; a_{\rm st} \; (T \; - \; T_0)  \; \; &  \; \; T_0  \; \lesssim \; T\;.
\end{array}
    \right.
\end{equation}

\begin{equation}
\label{6.18}
 w(T) \; \simeq \;  
 \left\{ \begin{array}{ll}
 \; \;   w(0)  \; \; &   \; \; T  \; \lesssim \; T_0  \;,
  \\
 \;  \; \frac{w(0)}{1 \; + \; a_{\rm w}  \; (T \; - \; T_0)}  \; \; &  \; \; T_0  \; \lesssim \; T \;,
\end{array}
    \right.
\end{equation}
taking $T_0 = 140$ MeV.

\begin{figure*}[htb]
\begin{center}
\includegraphics[width=0.49\linewidth,clip]{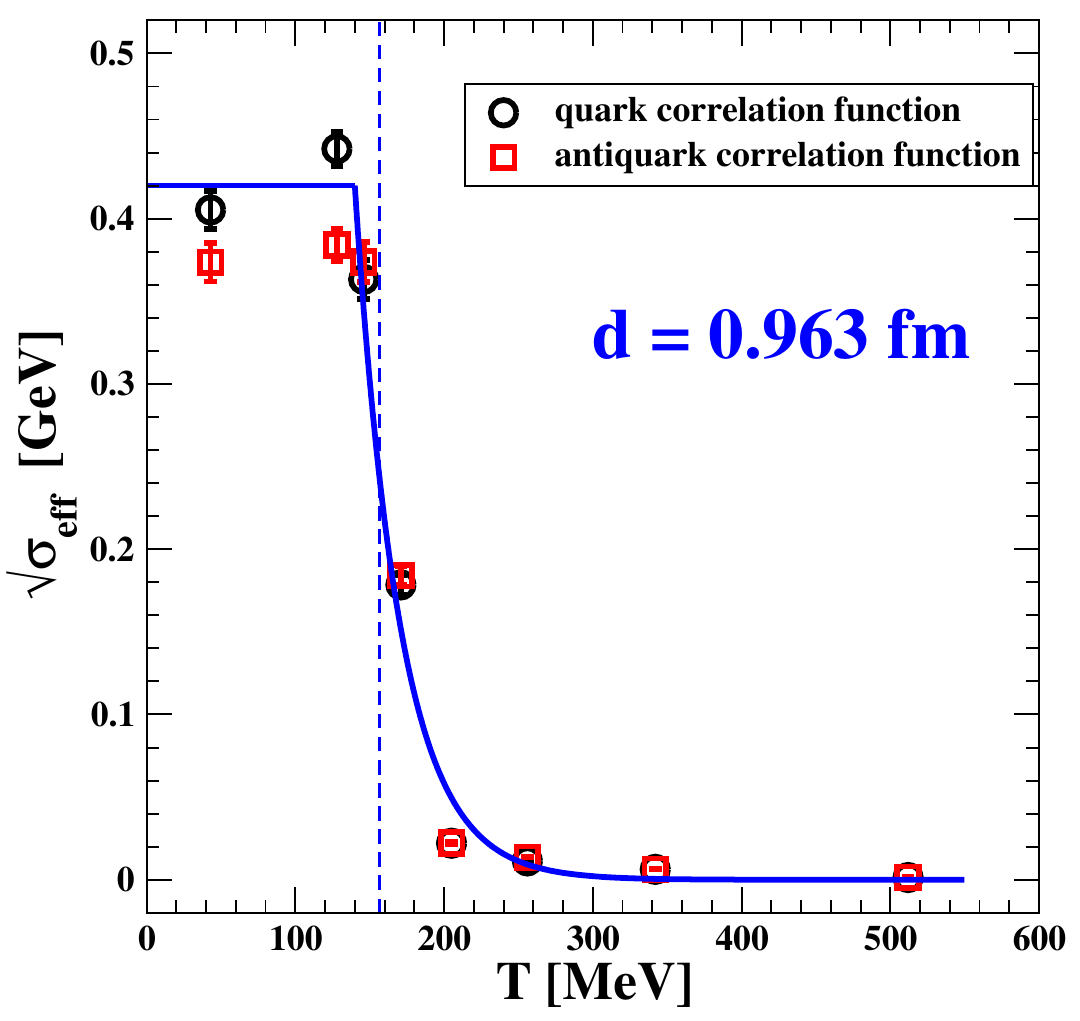}
\includegraphics[width=0.49\linewidth,clip]{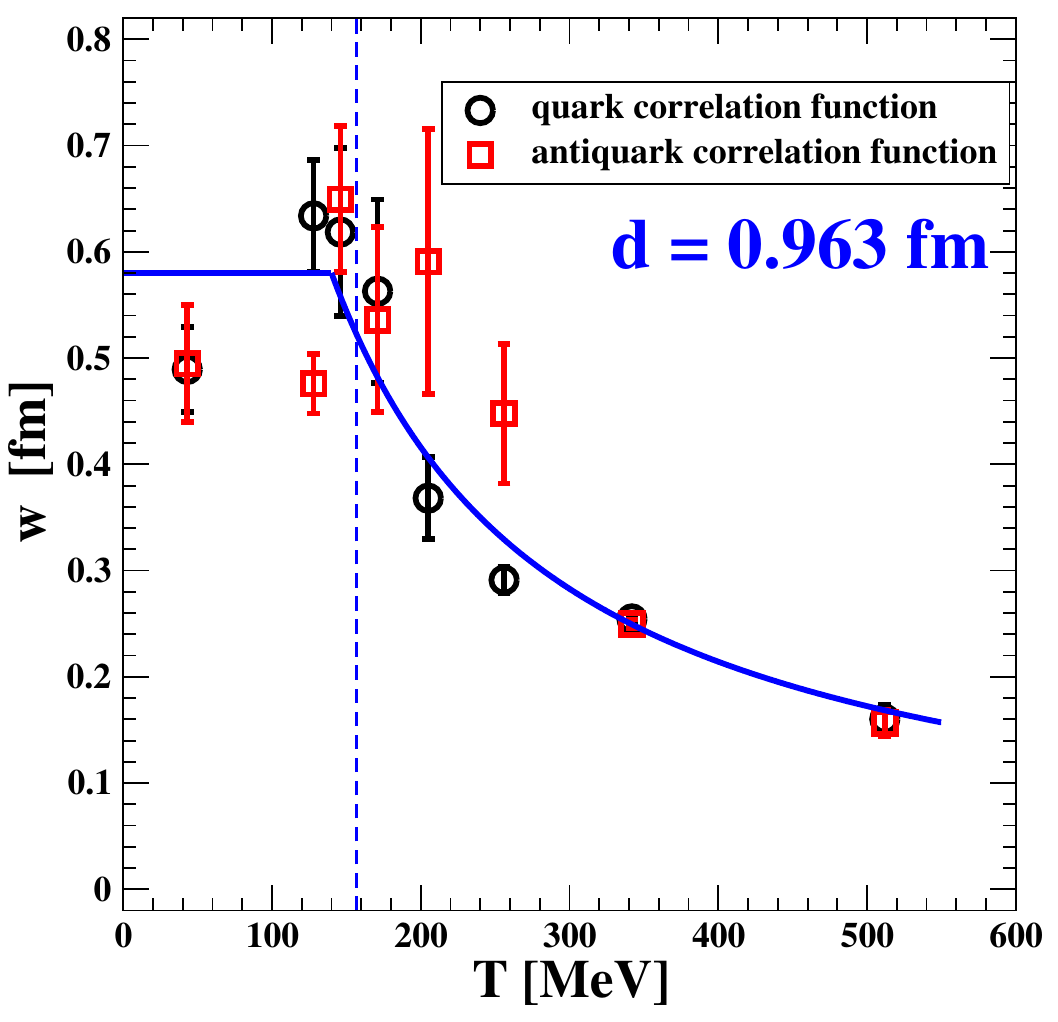}
\end{center}
\caption{(left) Effective string tension evaluated at the midplane for inter-quark distance
 $d \simeq$ 0.963 fm as a function of temperature, with best-fit to Eq.~(\ref{6.12}).
(right) Flux-tube width at the same distance, with best-fit to Eq.~(\ref{6.18}).}            
\label{fig:string-tension-width}
\end{figure*}

The collected data for effective string tension and width, together with the best fit curves are shown in Figure~\ref{fig:string-tension-width}.

\section{Summary}
\label{sec:summary}

We have studied the thermal behavior of the chromoelectric fields generated by a static quark-antiquark pair separated by
distances 0.57 fm $\lesssim  d \lesssim$  1.348 fm, for temperatures 
43  MeV $\lesssim  T \lesssim$  512 MeV. 

At all temperature values we see a narrow longitudinal field structure that can be identified as a flux tube, despite the largest considered temperature is more than three times larger than $T_\mathrm{pc} \approx 156$ MeV. 
The effective string tension remains almost constant from its zero temperature value to an onset temperature around 140 - 150 MeV. 
Above the onset temperature the string tension starts decreasing exponentially with $T$, while the flux-tube width decreases as $1/T$. 

This suggests that the flux tubes at $T > T_0$ are screened, with a screening mass growing with $T$, rather than fully disappear. 

\section*{Acknowledgments}
This investigation was in part based on the MILC collaboration's public lattice gauge theory code (\url{https://github.com/milc-qcd/}). Numerical calculations have been made possible through a CINECA-INFN agreement, providing access to HPC resources at CINECA. PC, LC and AP acknowledge support from INFN/NPQCD project. VC acknowledges support by  the Deutsche Forschungsgemeinschaft (DFG, German Research Foundation) through the CRC-TR 211  ``Strong-interaction matter under extreme conditions'' --  project number 315477589 -- TRR 211. This work is (partially) supported by ICSC – Centro Nazionale di Ricerca in High Performance Computing, Big Data and Quantum Computing, funded by European Union – NextGenerationEU.




\end{document}